\documentclass[doublecol]{epl2} 
\title{Vortex spectrum in superfluid turbulence:
interpretation of a recent experiment}
\shorttitle{Vortex spectrum in superfluid turbulence} %Insert here a short version of the title if it exceeds 70 characters

\author{P.-E. Roche\inst{1} \and C.F. Barenghi\inst{2} }
\shortauthor{P.-E. Roche \and C.F. Barenghi}

\institute{                    
  \inst{1} Institut N\'{E}EL, CNRS/UJF,
BP166, F-38042 Grenoble Cedex 9, France\\
  \inst{2} School of Mathematics, University of Newcastle,
  Newcastle upon Tyne NE1 7RU, UK
}

\abstract{We discuss a recent experiment in which the spectrum of the
vortex line density fluctuations has been measured in superfluid turbulence.
The observed frequency dependence of the spectrum, $f^{-5/3}$, disagrees with
classical vorticity spectra if, following the literature, the vortex line
density is interpreted as a measure of the vorticity or enstrophy. We argue that the
disagrement
is solved if the vortex line density field is decomposed into
a polarised field (which carries most of the energy) and an isotropic 
field (which is responsible for the spectrum).}

\newcommand{\bs}{\mathbf {s}}
\newcommand\bom{{\mbox{\boldmath $\omega$}}}

\begin{document}

\maketitle

%\begin{equation}
%\label{eq.1}
%0\neq1
%\end{equation}

%\begin{figure}
%\onefigure{epl-template.eps}
%\caption{Figure caption.}
%\label{fig.1}
%\end{figure}

\section{Motivation and aim}

Recent experiments have explored similarities and differences 
between turbulence in classical, ordinary fluids and
turbulence in He~II (superfluid turbulence). 
Superfluid turbulence consists of a tangle of quantised
vortex filaments; it is usually characterised (in both experiments
and numerical simulations) by the vortex line density $L$ (defined 
as the vortex
length per unit volume). Superfluid turbulence
 can be generated in many ways:
heat currents\cite{VanSciver-cylinder,VanSciver-counterflow,
Skrbek-counterflow}, 
vibrating wires\cite{Skrbek-fork}, oscillating grids\cite{Lancaster} or 
spheres\cite{Schoepe},  towed grids\cite{Stalp},  
bellows\cite{Smith1999,Fuzier2001},
rotating propellers \cite{Maurer1998,RocheEPL2007}
and  ultrasound\cite{Milliken1982}. Superfluid
turbulence is also tackled in the slightly different context of
superfluid $^3$He-B\cite{Helsinki,Fisher}. 

The current understanding of superfluid turbulence at the relatively
high temperature is the following. According to experimental 
\cite{Maurer1998,stalp2002}, theoretical \cite{Vinen2002}
and numerical \cite{KivotidesSpectra2002} 
results, at sufficiently large scales in the inertial 
range, the normal 
fluid and the superfluid components of He~II are strongly 
coupled, the superfluid and normal fluid velocities
are matched, $v_s \sim v_n$, and their energy
spectra obey the classical 
Kolmogorov law $k^{-5/3}$ (where the wavenumber $k$
and the frequency $f$ are related by $k=f/\overline{V}$ 
where $\overline{V}$ is the mean flow). 

In a recent experiment, Roche \etal \cite{RocheEPL2007}
measured the spectrum of the fluctuations of the vortex 
line density $L$ in turbulent superfluid helium $^4$He at
$T=1.6~\rm K$ and found a clear 
$f^{-5/3}$ dependence.
Our aim is to reconcile this observation with the 
current understanding of superfluid turbulence and
the interpretation (which is quite common
in the literature) of the vortex line density $L$ as a
measure of the superfluid vorticity, $\omega_s=\kappa L$,
where $\kappa \approx 10^{-7}\rm m^2/s$ is the 
quantum of circulation. Under this interpretation, the
$f^{-5/3}$ spectrum observed by Roche et al. \cite{RocheEPL2007} 
seems to contradict the scaling of vorticity observed 
in classical turbulence, which is a flat or slowly 
decreasing frequency spectrum 
(see for example \cite{Ishihara2003, Zhou2005} and 
references within).

To be more precise, what was actually measured in Ref. 
\cite{RocheEPL2007}
is the vortex line density corrected by a sine squared
prefactor to account for the orientation of each vortex 
line (as explained in Ref.\cite{Skrbek-counterflow} for example):
the component of a vortex line parallel to
the direction of sound propagation does not 
contribute to the second sound
attenuation at first order. The natural interpretation of the 
measured signal is therefore
the magnitude of the vorticity corrected by a prefactor 
calculated with the orientation of the vorticity vector. 
Using the DNS dataset\footnote{Data were downloaded from International Computational 
Fluid Dynamics database, hosted by the Cineca supercomputing center, 
Bologna, Italy. (http://cfd.cineca.it/)}
of Gotoh et al.\cite{Gotoh2002},
 we have checked that the correction introduced by this 
prefactor has only
a small contribution to the slope of the spectrum of the magnitude
of the vorticity in classical turbulence.
This still leaves us with a major discrepancy between the classical
vorticity spectrum and the 
steeper $f^{-5/3}$ decrease which was observed.

In bringing together the current understanding of 
superfluid turbulence
with the observed spectrum of $L$, we also need to make sure that the
interpretation of all measurements performed in 
Ref. \cite{RocheEPL2007} using a
pressure sensor and a second sound detector are consistent
with each other. In fact, at first sight there seems to 
be an inconsistency between the mean vortex line density and 
the energy of the flow 
estimated from the measured velocity

For a mean velocity 
$\overline{V} \approx 1\rm m/s$ at $T=1.6\rm ~K$, Roche
et al.\cite{RocheEPL2007} 
report a mean vortex line density $\overline{L}$
corresponding
to an average intervortex spacing $\delta$

\noindent
\begin{equation}
\label{eq-delta}
\delta = 1/\sqrt{\overline{L}} \approx 4 \times 10^{-6}\rm m.
\end{equation}

\noindent
from which we estimate backwards:
\begin{equation}
\label{eq-L}
\overline{L} = 1/{\delta^2} \approx 6 \times 10^{10}\rm m^{-2}.
\end{equation}

\noindent
The kinetic energy per unit volume
of the same flow is
\begin{equation}
\label{eq-K}
K=\frac{1}{2}\rho_n v_n^2 + \frac{1}{2} \rho_s v_s^2.
\end{equation}

\noindent
If the normal fluid and the superfluid are indeed coupled,
$v_n\approx v_s$, and if we assume that this velocity is 
approximately
$V_{rms}=0.3\rm m~s^{-1}$ (corresponding to the mean flow
$\overline{V}=1\rm m~s^{-1}$ and a turbulence intensity 
of $30 \%$), we have
\begin{equation}
\label{eq-K2}
K \approx \frac{1}{2} (\rho_n+\rho_s) V_{rms}^2
= \frac{\rho}{2} V_{rms}^2=6.5\rm J~m^{-3}.
\end{equation}

\noindent 
where $\rho=\rho_s+\rho_n=145\rm~kg~m^{-3}$.
Since $\rho_{n} < 0.2 \rho_{s}$, this energy $K$ is approximately
 equal to the superfluid kinetic energy. 
Let us assume that
the kinetic energy per
unit volume, $K$, is approximately equal to
the kinetic energy per unit length, $\cal E$, times
the length per unit volume $L$ :
\begin{equation}
\label{eq-K3}
K\approx{\cal E} L.
\end{equation}

\noindent
The kinetic energy per unit length is obtained in cylindrical
coordinates $(r,\phi,z)$ by integrating the 
square of the velocity field $\kappa/(2 \pi r)$ 
around a straight vortex line
(set along the $z$ direction) from the radial distance 
$r=a\approx 10^{-10}\rm m$ (the vortex core radius)
to some upper cutoff $b$:
\begin{equation}
\label{eq-E}
{\cal E}=\frac{\rho_s}{2} \int_0^{2 \pi} d\phi \int_a^{b}  r 
\left( \frac{\kappa}{2 \pi r}\right)^2 dr
=\frac{\rho_s \kappa^2}{4 \pi} \ln{(b/a)}.
\end{equation}

\noindent
If we take $b=\delta$, using
$\rho_s=122\rm~kg~m^{-3}$ at $T=1.6\rm K$, we have
\begin{equation}
\label{eq-E2}
{\cal E} \approx 1.0 \times 10^{-12}\rm J~m^{-1},
\end{equation}

\noindent
thus
\begin{equation}
\label{eq-L2}
L\approx K/{\cal E}=6 \times 10^{12}\rm m^{-2},
\end{equation}

\noindent
which is much bigger than the value of $\overline{L}$ 
from second sound measurements,
$L=6 \times 10^{10}\rm m^{-2}$. 
This second apparent inconsistency must be solved too. 

The following model which we propose
to solve these inconsistencies may not be the final answer.
Nevertheless, we think that
the exercise of putting together a coherent scenario with
the experimental information which is available at this stage is
a valuable exercise which should stimulate
further work and bring us closer to the correct solution of the
puzzle.

\section{Model}

The model which we propose to solve the puzzle described in the
previous section has two key features:
the decomposition of the vortex line distribution into
a ``polarised'' field and a ``unpolarised'' (or ``isotropic'')
field, and the assumption
that the unpolarised field has some statistical properties
of a passive vector field. We stress that our interpretation
is preliminary.

\subsection{Decomposition of the vortex line density}

Since the vortex core radius
is many orders of magnitude smaller than $\delta$ or
any other length scale of interest in the flow, 
we follow Schwarz \cite{Schwarz1988} and describe 
vortex lines as space curves ${\bf s}(\xi,t)$ where $\xi$ is arclength
and $t$ is time.
The quantity 
${\bf s}'=d{\bf s}/d\xi$ is the unit vector at the point
${\bf x}={\bf s}$ in the tangent
direction along the vortex line.

Consider a small cubic box $\Delta({\bf x})$
of size $\Delta > \delta$ and volume $\Delta^3$ centred around 
the point ${\bf x}$. 
We define the coarse--grained superfluid vorticity field as
\begin{equation}
\label{eq-omega}
\bom_s({\bf x})=\frac{\kappa}{\Delta^3} 
\int_{\Delta({\bf x})}{\bf s}' d\xi,
\end{equation}

\noindent
This definition
corresponds to the same coarse--graining procedure which
was used in Ref \cite{BarenghiModel1997} in a numerical calculation of a vortex 
tangle driven by an ABC normal flow to show that the (coarse--grained)
superfluid vorticity
matches the vorticity of the normal flow. 

Note that $\bom_s$ is nonzero only if the vortex lines are spatially 
organised. If the vortex lines point randomly in all directions,
then each Cartesian component of $\bom_s$ is zero, because in each
direction oriented vortex strands cancel each other out when summed
algebraically.

It is easy to check that the magnitude of $\bom_s$ is less
than $\kappa$ times the local vortex line density $L$: 
\begin{eqnarray}
\label{eq-omega2}
\vert \bom_s({\bf x}) \vert
=\vert \frac{\kappa}{\Delta^3} \int_{\Delta({\bf x})} {\bf s}' d\xi \vert 
\\ \nonumber
< \frac{\kappa}{\Delta^3} \int_{\Delta({\bf x})} 
\vert {\bf s}' \vert d\xi 
= \frac{\kappa}{\Delta^3} \int_{\Delta({\bf x})} d\xi=\kappa L({\bf x}),
\end{eqnarray}

\noindent
because $\vert {\bf s}'\vert =1$. 

From the coarse--grained superfluid vorticity we can
define the (local) polarised vortex line density
$L_{\parallel}$:
\begin{equation}
\label{eq-Lpar}
\kappa L_{\parallel}({\bf x})=\vert \bom_s({\bf x})\vert,
\end{equation}

\noindent
Since
\begin{equation}
\label{eq-Lpar2}
L_{\parallel}({\bf x}) < L({\bf x}).
\end{equation}
 
\noindent
the missing part is a field which we call $L_{\times}({\bf x})$ and we have
the (local) decomposition  
\begin{equation}
\label{eq-L3}
L({\bf x}) = L_{\times}({\bf x}) + L_{\parallel}({\bf x}) .
\end{equation}

%\noindent
%Integration is space yields

%\begin{equation}
%\label{eq-L4}
%L = L_{\times} + L_{\parallel}. 
%\end{equation}

The smoothed field $L_{\parallel}$ filters the vortex tangle
in $k$ space, getting rid of
short--wavelength Kelvin waves on the same vortex
line. It also accounts for cancellation effects arising from
vortex lines oriented in opposing directions. By construction, $L_{\parallel}$
is sufficiently organised that it defines
the coarse--grained superfluid vorticity field $\bom_s$. Thus
$L_{\parallel}$ reflects the superfluid
velocity field in the inertial range at scales larger than $\Delta$.
At scales smaller than $\Delta$ the superfluid vorticity field $L_{\parallel}$ is
clearly not defined.

Viceversa, $L_{\times}=L-L_{\parallel}$ contributes to the vortex line density 
but not to the superfluid
vorticity and reflects the randomly oriented vortex lines.
Note that we have not made any assumption
about the relative amount of wiggliness of $L_{\parallel}$ and 
$L_{\times}$.  Notice that $L_{\times}$ does not necessarily
consist only of small loops (left over by vortex reconnections
for example) or high energy Kelvin waves ($k\gg \Delta^{-1}$). 
Long filaments, provided they are randomly oriented with respect to 
their neighbours (so that they do not add
up vortex length in the same direction), can be part of $L_{\times}$.

\subsection{Passive vectors}

The second feature of our model is the assumption that the 
unpolarised field 
$L_{\times}$ has some statistical properties of
a classical passive vector field. 
It is well known \cite{Ohkitani} that passive vectors have a
power spectrum which obeys the $f^{-5/3}$ law, and this is our
explanation of the observed power spectrum of the vortex line
density.

At first it may seem contradictory to expect 
$L_{\times}$ to correspond to a
active field while assuming a passive nature for $L_{\parallel}$.
We now argue that this
active/passive distinction may result from a fundamental
property of superfluid vortices which makes them different from
classical vortices.

In a classical fluid \cite{ChorinLivre}
the local time derivative of the
vorticity is the combination of advection,
$( {\bf v} \cdot \nabla)\bom$, and stretching,
$(\bom \cdot \nabla){\bf v}$. An example
of the latter is the stretching of a classical vortex along its main axis
which elongates it while squeezing it transversally, resulting 
in an increase of vorticity. In the superfluid vortex stretching does 
not occur because the radius of the vortex core is fixed,
determined by quantum mechanical constraints on the rotation.
Superfluid vortices can become longer (for example, if energy is fed
from the normal fluid, or, at $T=0$, if the geometry changes keeping
the total kinetic energy constant), but their core is rigid, thus
they remain slender with respect to any typical turbulence
scale. Superfluid vortex dynamics thus differs from the dynamics
of classical vortices (for example see \cite{ChorinLivre,Zhou1997}).
Nevertheless, when superfluid vortices are assembled in polarised bundles,
the classical vorticity enhancement that is described above
can still be reproduced by stretching the whole bundle, as if
vortices were material lines. For example, axisymmetric
and non-axisymmetric oscillations of superfluid
vortex bundles in the form of waves are known in the literature
\cite{HendersonBarenghi}.
It is therefore reasonable to expect
that the field $\bom_s$, which result from polarised vortices, will
mimic a classical active vorticity field at large enough scales.

We now turn to the stretching of an \textit{unpolarised} tangle
by a large scale velocity field. If, as modelled above, 
superfluid vortices behave as material
lines, it is also reasonable to expect that
the tangle will remain unpolarised, and, due to the fluid 
incompressibility, the total
length of lines will remain unchanged : in other words,
$L_{\parallel}$  will be
simply transported by the coarse-grained velocity.

Viceversa, it is reasonable to assume
that the high density and large density fluctuations of $L_{\times}$
have little impact on the dynamics of the polarised field  $L_{\parallel}$ or $\bom_s$,
that is to say that $L{\times}$ does not advect $L_{\parallel}$
at scale large than $\Delta$ (where $L_{\parallel}$
is defined). This must be the case, because the velocity field 
induced by the unpolarised field must be very short-ranged,
probably $1/r^2$, caused by multipolar sources with no contribution 
at first order $1/r$.

\section{Consistency with measurements}

\subsection{The spectrum of the polarised field}

Our model is consistent only if we can show that the polarised
vortex line density $L_{\parallel}$ gives a negligible
contribution to the spectrum below $1\rm kHz$, which is the observed
second sound frequency range used by Roche \etal.

Following what said in the introduction, we
approximate the second sound spectrum $P_{\parallel}$ 
arising from $L_{\parallel}$ as
a white noise signal up to a viscous
cutoff corresponding to few times the frequency
of the Kolmogorov length scale $\eta$:
\begin{equation}
\label{eq-Ppar}
P_{\parallel} = \frac{\overline{L_{\|} ^2}-\overline{L_{\|}}^2}
{\overline{V}/ (4\eta) } 
\end{equation}

The denominator is the full frequency span of an ideal 
second sound probe. Eq.\ref{eq-Ppar}
represents the (constant) power spectral density at all frequencies
below the cut-off frequency. An ideal second sound probe
is fixed in space. The 
smallest time scales which are visible to the probe are produced
by the smallest flow structures (of typical size $4 \eta$)
which are advected past the probe at the local fluid velocity
(approximated by the mean flow velocity). The inverse of this time 
scale gives the highest frequency of the signal seen by the ideal probe,
which is indeed the frequency span.

Let us find an upper bound for $P_{\parallel}$. 
In classical turbulence, both experimental \cite{Antonia1998}
and numerical studies
%(as we checked using the data of Gotoh \etal\cite{Gotoh2002})
suggest that the vorticity $\bom$ roughly satisfies

\begin{equation}
\overline{\omega^2}\sim 2\overline{ | \omega | }^2,
\end{equation}

If we assume that this classical relation applies to the the normal 
fluid in the experiment \cite{RocheEPL2007}, and if we make the 
further assumptions
that normal fluid and superfluid are locked, $\omega_s \approx
\omega_n$, and that
$\kappa L_{\parallel}=\omega_s$, we have
\begin{equation}
\overline{L_{\parallel} ^2} \simeq 2\overline{L_{\parallel}}^2
\end{equation}

Thus
\begin{equation}
P_{\parallel} \sim 
\frac{2 \eta}{\overline{V}} {\overline{L_{\parallel}^2}}
= \frac{2 \eta}{\overline{V}} 
\left(\frac{\mu \kappa^2 {\overline{L_{\parallel}^2}}}
{\mu \kappa^2} \right),
\end{equation}

\noindent
where $\nu=\mu / \rho$ is the kinematic viscosity and $\mu$ the 
viscosity of helium.

A bound for $P_{\parallel}$ can be found by noticing that 
the total rate of dissipation of kinetic
energy per unit volume in turbulent He~II, $\rho \epsilon$,
should be larger than the dissipation
$\mu \omega_{n}^2$ which arises
from the regular viscous dissipation in the normal fluid alone:  
\begin{equation}
 \mu \omega_{n}^2 < \rho \epsilon,
\end{equation}

\noindent
where,  since $\omega_s \approx \omega_n$,
\begin{equation}
\mu \omega_{n}^2 \approx \mu \omega_{s}^2 = 
\mu \kappa^2 \overline{L_{\parallel}^2 },
\end{equation}

\noindent
Using the kinematic viscosity $\nu=\mu / \rho$, we find
\begin{equation}
P_{\parallel} <  \frac{2 \eta}{\overline{V}} 
\left(\frac{\epsilon} {\nu \kappa^2} \right),
\end{equation}

\noindent
In the next section, we present different ways of evaluating
the Kolmogorov length $\eta$ at $T=1.6K$ and show that they all give 
the same order of magnitude for $\eta$. For clarity, we use the
following classical expression (with the kinematic viscosity 
defined above):
\begin{equation}
\eta \approx (\nu^3/\epsilon)^{1/4}.
\end{equation}

\noindent
We have
\begin{equation}
P_{\parallel} <  \frac{2}{\kappa^2 \overline{V}} 
\left( \frac{\epsilon^3}{\nu} \right)^{1/4},
\end{equation}

To evaluate this expression, we estimate the rate of kinetic energy
dissipation $\epsilon$ at the integral scale,
$\ell_0 \approx 10^{-2}\rm m$, for which 
$V_0\approx V_{rms}=0.3\rm m/s$,
and obtain
\begin{equation}
\epsilon \approx \frac{V_0^3}{\ell_0} 
\approx \frac{V_{rms}^3}{\ell_0}
\approx 2.7\rm m^2~s^{-3},
\end{equation}

At $T=1.6\rm K$, $\nu \approx 8.9 \times 10^{-9}\rm m^2~s^{-1}$ 
and we obtain
$P_{\parallel} < 4.3 \times 10^{16}\rm m^{-4}~s$, 
which is much less then the observed spectral density
$P=2 \times 10^{19}\rm m^{-4}~s$ and is just above
the instrumental noise level $0.5 \times 10^{16}\rm m^{-4}~s$
(see Fig 4 of Ref. \cite{RocheEPL2007}).
We conclude that the contribution of $L_{\parallel}$ to the
observed spectrum is negligible. 

\subsection{The Kolmogorov length}

We estimate the Kolmogorov length from the expression
$\eta=(\nu^{\star 3}/\epsilon)^{1/4}$.
In principle we can define three possible 
kinematic viscosities $\nu^{\star}$ in our problem
(all numerical values refer to $T=1.6\rm K$). The first is based on
the total density $\rho=\rho_n+\rho_s=145\rm kg~m^{-3}$ and is
$\nu=\mu/\rho=8.9 \times 10^{-9}\rm m^2~s^{-1}$. The second is based on
the normal fluid density $\rho_n=23.6\rm kg~m^{-3}$ and is
$\nu_n=\mu/\rho_n=55 \times 10^{-9}\rm m^2~s^{-1}$. The third
is the efficient kinematic viscosity 
$\nu'\approx 20 \times 10^{-9}\rm m^2~s^{-1}$ determined from
towed--grid experiments in turbulent He~II, see fig.8 of Ref .\cite{stalp2002}
Using these three values, we obtain respectively
$\eta \approx 0.7\rm \mu m$,
$\eta \approx 3\rm \mu m$ and
$\eta \approx 1\rm \mu m$. In all cases $\eta$ is 
of the order of magnitude of the intervortex spacing,
$\delta \approx 4\rm \mu m$ or at the most six times smaller.
This is consistent with the argument of Vinen and Niemela \cite{Vinen2002} that 
$\delta$ and $\eta$ are likely to be of the same order of magnitude:
the superfluid and the normal fluid are coupled throughout the
inertial range.

\subsection{The energy}

A key feature of our model is that the unpolarised field $L_{\times}$
gives a negligible
contribution to the total energy of the flow compared to
the polarised field $L_{\parallel}$. The  estimate made in the
introduction that
the kinetic energy per unit volume is
$K\approx 6.5\rm J m^{-3}$
clearly refers to the polarised field $L_{\parallel}$, 
because we obtained
it using the condition $v_n\approx v_s$: in this notation we rewrite
\begin{eqnarray}
K_{\parallel}=\frac{1}{2} \rho_n v_n^2 + \frac{1}{2} \rho_s v_s^2
\approx \frac{1}{2} (\rho_n+\rho_s) V_{rms}^2\\ \nonumber
= \frac{\rho}{2} V_{rms}^2=6.5\rm J~m^{-3}
\end{eqnarray}

Let us estimate the energy contained in the unpolarised field $L_{\times}$.
If we picture $L_{\times}$ as a random network of straight vortex lines,
the energy per unit volume can be obtained following the
integral procedure that leads to Eq.\ref{eq-E} with $b=\delta$: we
multiply the length per unit volume times the integral
of the square of the velocity field 
only up to a radial distance
which is of the order of the intervortex spacing, because
at this distance the total velocity fields of randomly oriented
vortices cancel each other out.
Using the numerical value provided in Eq.\ref{eq-E} and 
Eq.\ref{eq-L} for $L_{\times} \approx \overline{L}$, we get
\begin{equation}
K_{\times} \approx 
\frac{\rho_s \kappa^2}{4 \pi} L_{\times}\ln{(\delta/a)}
\approx 0.06~J m^{-3} << K_{\parallel}.
\end{equation}

\noindent
The above integration procedure would not give the kinetic energy per unit volume arising
from the polarised field $L_{\parallel}$, because contributions of 
different vortex lines add up rather than cancel each other.
This why we expect $K_{\parallel} >> K_{\times}$
to hold when a sufficient level of vortex polarisation is reached.

The argument is made more clear if we
consider a cylindrical container of radius $h$ and height $h$
containing $N$ straight vortex lines aligned along the axis.
The number of vortex lines per unit area is
$L=N/(\pi h^2)$ and the intervortex distance is
$\delta=\sqrt{\pi h^2/N}$.  Suppose that the vortices are oriented
in the same direction (complete polarisation), forming a vortex
bundle,
as in a recent numerical calculation
\cite{Kivotides2007}; then the velocity fields of the vortices
add up and create a total solid--body rotation velocity $v=\Omega r$ where
$\Omega$ is obtained from
\begin{equation}
N \kappa= \oint_C{\bf v} \cdot {\bf d\ell} 
= \int_S \nabla \times {\bf v} \cdot {\bf dS}
=\int_S \bom \cdot {\bf dS} =2\Omega \pi h^2,
\end{equation}

\noindent
which yields

\begin{equation}
\Omega=N\kappa/(2 \pi h^2).
\end{equation}

\noindent
The energy per unit volume is
\begin{equation}
E_{\parallel}=\frac{\rho_s \kappa^2 h^2 L^2}{16},
\end{equation}

Now assume the opposite limit, that the vortices are randomly oriented 
in the positive or negative direction along the axis of the cylinder.  
The energy per unit volume is
\begin{eqnarray}
E_{\times}=\frac{1}{\pi h^3}\frac{\rho_s}{2} N \int_0^h dz \int_0^{2 \pi} d\phi
\int_a^{\delta} dr r \left(\frac{\kappa}{2 \pi r}\right)^2\\ \nonumber
=\frac{\rho_s\kappa^2 L}{8 \pi} \ln{(1/(L a^2))},
\end{eqnarray}

\noindent
(We write it in terms of $L$ rather than $N$ because we want to take
the limit of increasing the density rather than the number of vortices).
Therefore
\begin{equation}
\frac{E_{\times}}{E_{\parallel}}
=\frac{2}{\pi} \frac{1}{L h^2}\ln{(1/(L a^2))}
=\frac{4}{\pi} \frac{\delta^2}{h^2}\ln{(\delta/a}),
\end{equation}

Clearly, the larger is the vortex line density $L$, the smaller is
$E_{\times}$ with respect to $E_{\parallel}$ (provided $\delta>a$ of course).

\section{Conclusion}

Our model of superfluid turbulence 
consists in dividing the tangle in a polarised field 
$L_{\parallel}$ and a unpolarised field $L_{\times}$ such that the
total vortex line density is $L=L_{\parallel}+L_{\times}$.
The polarised field makes up the (coarse--grained)
superfluid vorticity field, such that the two fluids are coupled
($v_n \sim v_s$) in the
inertial range according to current understanding of superfluid
turbulence.
The unpolarised field has some of the statistical properties of
a passive vector field.
What is observed with the second sound probe is mainly $L_{\times}$,
not $L_{\parallel}$, which, as we have shown, brings a negligible
contribution to the measured spectrum: 
\begin{equation}
P_{\times}>>P_{\parallel}.
\end{equation}

\noindent
However the polarised field $L_{\parallel}$ has more
energy than $L_{\times}$:
\begin{equation}
K_{\parallel}>>K_{\times}
\end{equation}

\noindent
That is why if we try to infer the
vortex line density from Eq.\ref{eq-L2} 
we get an unrealistic high value.

The model suggests the following picture of the turbulent tangle 
in the large scales (upper inertial range): large fluctuations of 
the density of vortex line on top of a small modulation of 
polarisation which -nevertheless- controls the velocity field, 
energy cascade
and passive advection of most of the vortex line density.

 We stress that the model which we propose is only an attempt
to combine the information which is available from the experiment
of Roche \etal \cite{RocheEPL2007} in a consistent scenario.
If the model can be confirmed, 
the  $f^{-5/3}$ power law dependence of the fluctuations of
the vortex line density reported in \cite{RocheEPL2007}
 should be considered as the
inertial-range signature of the quantum
nature of superfluid turbulence.

Finally we remark that
a decomposition of the vortex line density field
in polarised and isotropic parts similar in spirit 
to what we have done, has been
attempted by Lipniacki\cite{Lipniacki}. Lipniacki's theory results in 
an Euler equation (motified by the presence of friction) which is similar
to Hall--Vinen equation for the macroscopic superfluid velocity 
\cite{Henderson}, which is coupled to a modified Vinen equation for the
(more microscopic and isotropic) vortex line density $L$. 

\acknowledgments
C.F.B. is supported by EPSRC grants GR/T08876/01 and EP/D040892/1
and P.-E. R. by the ANR grant TSF.


\begin{thebibliography}{0}

\bibitem{VanSciver-cylinder}
\Name{Zhang T. \and Van Sciver S.}
\REVIEW{Nature Physics}{1}{36}{2005}

\bibitem{VanSciver-counterflow}
\Name{Zhang T. \and Van Sciver S.}
\REVIEW{J. Low Temp. Physics}{138}{865}{2005}

\bibitem{Skrbek-counterflow}
\Name{Barenghi C.F., Gordeev A.V. \and Skrbek L.}
\REVIEW{Phys. Rev. E}{74}{026309}{2006}

\bibitem{Skrbek-fork}
\Name{Blazkova M., Schmoranzer D \and Skrbek L.}
\REVIEW{Phys. Rev. E}{75}{025302}{2007}

\bibitem{Lancaster}
\Name{Charambolous D., Skrbek L, Hendry P.C.,McClintock P.V.E.
\and Vinen W.F.}
\REVIEW{Phys. Rev. E}{74}{036307}{2006}

\bibitem{Schoepe}
\Name{Jager J., Schuderer B. \and Schoepe W.}
\REVIEW{Phys. Rev. Lett.}{74}{566}{1995}

\bibitem{Stalp} 
\Name{Stalp S.R., Skrbek L. \and Donnelly R.J.}
\REVIEW{Phys. Rev. Lett.}{82}{4831}{1999}

\bibitem{Fuzier2001}
%Steady-state pressure drop and heat transfer in {He II} forced flow at high {R}eynolds number
\Name{Fuzier S., Baudouy B. \and Van Sciver S. W.}
\REVIEW{Cryogenics}{41}{2001}{453}.

\bibitem{Smith1999}
%Observed drag crisis on a sphere in flowing {H}e {I} and {H}e {II}}
\Name{Smith M.R. , Hilton D.K. \and Van Sciver S. W.}
\REVIEW{Phys. Fluids}{11}{1999}{751}.

\bibitem{Maurer1998}
\Name{Maurer J. \and Tabeling P.}
% Local investigation of superfluid turbulence.
\REVIEW{Europhysics Lett.} {43}{1998}{29}.

\bibitem{RocheEPL2007}
\Name{Roche P.-E., Diribarne P., Didelot T., Fran{\c c}ais O.,
Rousseau L. \and Willaime H.}
% Vortex density spectrum of quantum turbulence.
\REVIEW{EPL} {77}{2007}{66002}.

\bibitem{Milliken1982}
\Name{Milliken F.P., Schwarz K.W. \and Smith C.W.}
%Free decay of superfluid turbulence.
\REVIEW{Phys. Rev. Lett.}{48}{1982}{1204}.

\bibitem{Helsinki}
\Name{Finne A.P., Araki T., Blaauwgeers R.,
Eltsov V.B., Kopnin N.B., Krusius M., Skrbek L., Tsubota M. \and Volovik G.E.}
\REVIEW{Nature}{424}{1022}{2003}

\bibitem{Fisher}
\Name{Bradley D.I., Clubb D.O., Fisher S.N.,
Guenault A.M., Haley R.P., Matthews C.J., Pickett G.R., Tsepelin V.
\and Zaki K.}
\REVIEW{Phys. Rev. Lett.}{93}{035302}{2005}

\bibitem{stalp2002}
\Name{Stalp S. Niemela J.J., Vinen W.F., \and  Donnelly R.J.}
% Dissipation of grid turbulence in helium II
\REVIEW{Phys. Fluids} {14}{2002}{1377}.

\bibitem{Vinen2002}
\Name{Vinen W.F. \and Niemela J.J.}
% Quantum turbulence.
\REVIEW{J. Low Temp. Phys.}{128}{2002}{128}.

\bibitem{KivotidesSpectra2002}
\Name{Kivotides D., Vassilicos J.C., Samuels D.C. \and Barenghi C.F.}
% Velocity spectra of superfluid turbulence.
\REVIEW{Europhys. Lett.}{57}{2002}{845}.

\bibitem{Ishihara2003}
\Name{Ishihara T., Kaneda Y., Yokokawa M., Itakura K. \and Uno A.}
% Spectra of energy dissipation, enstrophy and pressure by
% high-resolution direct numerical simulations of turbulence 
%in a periodic box.
\REVIEW{J. Phys. Soc. Japan} {72}{2003}{983}.

\bibitem{Zhou2005}
\Name{Zhou T., Antonia R.A. \and  Chua L.P.}
% Flow and Reynolds number dependencies of one-dimensional 
% vorticity fluctuations.
\REVIEW{J. of Turbulence}{6}{2005}{28}.


\bibitem{Gotoh2002}
\Name{Gotoh T., Fukayama D. \and Nakano T.}
% Velocity field statistics in homogeneous steady turbulence obtained
% using a high-resolution direct numerical simulation.
\REVIEW{Phys. Fluids}{14}{2002}{1065}.


\bibitem{Schwarz1988}
\Name{Schwarz K.W.}
% 3-dimensional vortex dynamics in superfluid he-4 - homogeneous
% superfluid turbulence.
\REVIEW{Phys. Rev. B}{38}{1988}{2398}.

\bibitem{BarenghiModel1997}
\Name{Barenghi C.F., Samuels D.C.  \and  Bauer G.H.}
% Superfluid vortex lines in a model of turbulent flow.
\REVIEW{Phys. Fluids}{9}{1997}{2631}.

\bibitem{Antonia1998}
\Name{Antonia R.A., Zhou T. \and Zhou Y.}
% Three-component vorticity measurements in a turbulent grid flow.
\REVIEW{J. Fluid Mech.}{374}{1998}{29}.

\bibitem{Ohkitani}
\Name{Ohkitani K.}
\REVIEW{Phys. Rev. E}{65}{2002}{046304}.

\bibitem{ChorinLivre}
\Name{ Chorin, A.}
\REVIEW{Vorticity and Turbulence}{}{1997}{Springer 2nd ed.}.

\bibitem{Zhou1997}
%On the motion of slender vortex filaments
\Name{Zhou, H.}
\REVIEW{Phys. Fluids}{9}{1997}{970}.

\bibitem{HendersonBarenghi}
\Name{Henderson K.L. and Barenghi C.F.}
% Vortex waves in a rotating superfluid
\REVIEW{Europhys Lett.} {67}{2004}{56}.

\bibitem{Kivotides2007}
\Name{Kivotides D.}
% Coherent structure formation in turbulent thermal superfluids
\REVIEW{Phys. Rev. Lett.}{96}{2007}{175301}.

\bibitem{Lipniacki}
\Name{Lipniacki T.}
%Dynamics of superfluid He-4: Two-scale approach 
\REVIEW{European J. Mech. B-Fluids}{25}{2006}{435}

\bibitem{Henderson}
%Nonlinear Taylor - Couette flow of helium II
\Name{Henderson K.L., Barenghi C.F. \and Jones C.A.}
\REVIEW{J. Fluid Mech.}{283}{1995}{329}.

\end{thebibliography}
\end{document}